\documentclass[submitted]{IEEEtran}
\usepackage{amsmath}
\usepackage{soul}
\usepackage{siunitx}
\usepackage[table]{xcolor}
\usepackage{cite}  
\usepackage{multirow}
\usepackage{makecell}
\usepackage{color}
\usepackage[colorlinks=true,linkcolor=blue,urlcolor=blue,citecolor=blue]{hyperref}
\usepackage[colorinlistoftodos]{todonotes}

\begin{document}


\title{Compressive Sensing Empirical Wavelet Transform for Frequency-Banded Power Measurement\\ Considering Interharmonics}

\author{Jian Liu, Wei Zhao, and Shisong Li$^\dagger$, \textit{Senior Member, IEEE}
\thanks{Jian Liu, Wei Zhao, and Shisong Li are with the Department of Electrical Engineering, Tsinghua University, Beijing 100084, China. Wei Zhao is also with the Yangtze Delta Region Institute of Tsinghua University, Jiaxing, Zhejiang 314006, China.}
\thanks{This work was supported by the Smart Grid-National Science and Technology Major Project under Grant No. 2024ZD0803300.}
\thanks{$^\dagger$Email: shisongli@tsinghua.edu.cn}}

\maketitle

\begin{abstract}
Power measurement algorithms based on Fourier transform are susceptible to errors caused by interharmonics, while wavelet transform algorithms are particularly sensitive to even harmonics due to band decomposition effects. The empirical wavelet transform (EWT) has been demonstrated to improve measurement accuracy by effectively partitioning transition bands. However, for detecting interharmonic components, the limitation of the observation time window restricts spectral resolution, thereby limiting measurement accuracy. To address this challenge, this paper proposes a Compressive Sensing Empirical Wavelet Transform (CSEWT). The approach aims to enhance frequency resolution by integrating compressive sensing with the EWT, allowing precise identification of components across different frequency bands. This enables accurate determination of the power associated with the fundamental frequency, harmonics, and interharmonics. Test results indicate that the proposed CSEWT method can significantly improve the precision of individual frequency component measurements, even under dynamic and noisy conditions.
\end{abstract}


\section{Introduction}
\IEEEPARstart{I}{n modern} power systems, electrical signals, particularly current signals, often deviate significantly from ideal sinusoids, displaying notable fluctuations in both the time and frequency domains. From a signal processing perspective, these signals consist of several components: the fundamental frequency, harmonics, interharmonics, superharmonics, and noise. Conventional power metering studies, guided by the IEEE 1459-2010 standard \cite{1}, have predominantly focused on the fundamental frequency and harmonics. However, with the growing integration of renewable energy sources, such as wind and solar power, and the increasing use of large-scale dynamic loads, such as electric vehicles, the presence of signal components beyond the fundamental and harmonics—especially interharmonics—has become more prevalent in power grids. Interharmonics, which are non-integer multiples of the fundamental frequency, introduce complexity into power flow analysis because they are no longer orthogonal to the fundamental or harmonic components. As a result, signal measurement methods and associated analytical algorithms must account for the influence of interharmonics to ensure accurate monitoring and control of modern power systems.

In 2007, the IEEE Task Force on Harmonics Modeling and Simulation investigated the generation mechanisms of interharmonic sources in power grids, from fundamental concepts to models for various types of interharmonic \cite{2}. In \cite{3}, based on an inherent relationship between voltage fluctuation and interharmonic, Jing Yong \textit{et al}, demonstrated that an equivalent single interharmonic component could be used to represent the voltage fluctuation effect of a pair of interharmonics. To solve the flicker-detection problem associated with interharmonics, Taekhyun Kim \textit{et al} proposed a new approach based on down-up sampling. The method extends the scope of interharmonic-related flicker detection and complements the IEC standard \cite{4}. The generation mechanisms of variable frequency interharmonic currents from the inverter to the rectifier were investigated in \cite{5}. Duro Basic determined that, as the drive output frequency changes, these interharmonics can generate between the DC and the third harmonic randomly. In \cite{6}, the modified harmonic domain (MHD) and the modified dynamic harmonic domain (MDHD) were enhanced to represent interharmonics, allowing the model of elements of the power system considering interharmonics. It was demonstrated that not all of the interharmonics identified in the DFT analysis were genuine. In order to determine the existence of genuine interharmonics, a time–frequency contour analysis was proposed in \cite{7} to reveal frequency characteristics over a period of time. In \cite{8}, a new decomposition method for nonsinusoidal and unbalanced conditions was proposed to address the the unbalance evaluation, and the method extends the IEEE Standard 1459-2010, taking the interharmonic into account. In \cite{9}, Ariya Sangwongwanich \textit{et al} explore the characteristics of interharmonics in PV systems, and propose a model of interharmonics. The model can predict the frequencies and amplitudes of interharmonics based on the parameters of the maximum power point tracking (MPPT) algorithm.

{Interharmonics cause distortion in voltage and current signals, especially in current signals, leading to the inaccurate energy consumption and the unfair electricity bills. Additionally, signals may contain interharmonics, having, e.g., amplitude modulated signals, introducing extra power components that disrupt grid power metering \cite{10}.} Therefore, in electrical power measurement, a key challenge is adapting power metering algorithms to accurately analyze signals containing interharmonics. In general, power metering methods are broadly classified into four categories: time-domain, frequency-domain, time-frequency-domain, and parametric methods. Among these, time-domain methods, such as the widely used dot product sum algorithm, are limited in their ability to differentiate between harmonic and interharmonic power flows. In contrast, methods based on the frequency-domain, time-frequency-domain, and parametric approaches have been developed for measuring interharmonics \cite{11}.

In the frequency domain, the DFT is a widely used method for harmonic power analysis.  To address issues of aliasing effect, spectral leakage, and fence effect, a recursive group-harmonic power minimizing algorithm is proposed in \cite{12} for harmonic and interharmonic analysis, significantly improving signal identification accuracy compared to DFT. 
In \cite{13}, Carlos M. Orallo \textit{et al} employ the sliding DFT (SDFT) and variable sampling period technique (VSPT) to accurately determine the harmonics, with strong robustness and frequency adaptability.
To solve the spectrum interference of multiple adjacent interharmonics under asynchronous sampling condition, a multi-interharmonic spectrum separation and measurement based on DFT, is proposed in \cite{14} to detect and estimate intensive interharmonics, accurately identifying the frequency components and parameters.
Kai Wang \textit{et al} employ the Chirp-Z transform (CZT) \cite{15} and interpolation discrete Fourier transform (IpDFT) \cite{16} in sequence, established the linear equation set to describe the correspondence between the frequency, amplitude, phase and observed spectrum, fully utilizing the information of side-lobe interference and spectrum leakage.
{Furthermore, Kitzig \textit{et al.} propose a fast-decaying sine ramp window approach to reduce far-distant spectral leakage in harmonic analysis, achieving enhanced frequency resolution and measurement accuracy under dynamic conditions \cite{17}.}
These methods analyze interharmonics through analogy, suppression, separation, and reconstruction based on spectral information, offering valuable insights for interharmonic analysis. However, their application to power measurements containing interharmonics has not yet been demonstrated.

In parametric methods, an estimation of signal parameters via rotational invariance technique (ESPRIT)-based method is proposed in \cite{18} for harmonics and interharmonics detection, estimating frequency, amplitude and phase accurately with strong noise immunity and robustness. 
Based on the subspace and least mean square,  an S-LMS method is developed in \cite{19}, and enhanced in \cite{20}, to detect the harmonics and interharmonics. The method boasts high-frequency resolution, improved noise thresholds, and immunity to fundamental frequency deviations.
In \cite{21}, Babak Jafarpisheh \textit{et al} put forth an adaptive accelerated MUSIC (A$^{2}$MUSIC) algorithm, that enhances the precision of frequency estimation while minimizing the computational burden and exhibiting resilience to noise.
In order to prevent the spectral interference produced by interharmonic, Jian Song \textit{et al} proposed Taylor weighted least square matrix pencil (TWLSMP) in \cite{22}, capturing time-varying characteristics of interharmonics.
In \cite{23}, N. A. Yalcin and F. Vatansever integrate the Prony method with Haar transform coefficients, thereby facilitating the accurate and expeditious calculation of interharmonic parameters with strong robustness.
These parameterized methods are capable of precisely estimating interharmonic frequency parameters while exhibiting high noise immunity. However, the lengthy processing time and the necessity for preprocessing, in some cases, restrict its applicability in power metering.

In the time-frequency-domain, time-frequency distribution (TFD) and cross-time-frequency distribution (XTFD) are used to evaluate the instantaneous power components, and the method proposed in \cite{24} overcomes the limitations of periodic signals. 
To reconstruct estimated noiseless signal, the joint method of the singular value decomposition (SVD) and the short-time Fourier transform (STFT) is proposed in \cite{25}, which provides the de-noising of Partial Discharge (PD) signals.
Walid G.Morsi and Mohamed E.El-Hawary adapt the wavelet packet transform (WPT) to power expressions in the IEEE standard 1459-2010, extending the power definition of interharmonic components and preserving both time-frequency information while improving measurement accuracy \cite{26,27}.
To separate interharmonic components, the undecimated wavelet packet transform (UWPT) was designed in \cite{28}, with the Hilbert transform (HT) applied to calculate the instantaneous amplitude and phase of each frequency, enabling effective monitoring of different signal parameters. 
In order to provide accurate estimations for non-stationary signals, the Stockwell Transform (ST) has been applied to measuring the frequency, magnitude and phase of signals in \cite{29}, extending the scope of the IEEE standard 1459-2010.

The time-frequency domain method retains time information for each frequency component, allowing for more effective analysis of the time-varying characteristics of interharmonics. This approach is increasingly being applied in electrical power measurement, although challenges remain in enhancing frequency resolution. IEC 61000-4-7 recommends an observation window length of 10 fundamental cycles, corresponding to a spectral resolution of 5\,Hz \cite{30}. This standard provides essential guidance for the analysis of both harmonics and interharmonics. However, the frequency-domain and time-domain methods are constrained by this 5\,Hz spectral resolution. 
{While parametric methods can achieve higher spectral resolution compared to non-parametric approaches, their performance is fundamentally constrained by the Cramér–Rao Bound (CRB), which makes it challenging to achieve precise spectral resolution below 5\,Hz. This limitation becomes particularly critical in power systems where interharmonics do not align as integer multiples of 5\,Hz. To accurately resolve such closely spaced frequencies, it is essential to minimize the variance of the estimators. The CRB reveals that frequency estimation precision is heavily dependent on factors such as signal amplitude, noise levels, data length, and the proximity of spectral components. In practice, when these interharmonics are closely spaced or exist in low signal-to-noise ratio (SNR) conditions, even advanced parametric methods struggle to approach the theoretical limits set by the CRB.}

In such cases, all three methods—frequency-domain, time-frequency-domain, and parametric—encounter difficulties in accurately measuring these interharmonic parameters due to insufficient spectral resolution.
To address this issue, we presented a compressive sensing empirical wavelet transform(CSEWT) for precise power measurement that accounts for interharmonics at the 2024 Conference on Precision Electromagnetic Measurements (CPEM 2024, Denver, USA)~\cite{31}. This paper is an extended version of \cite{31}, in both theoretical and experimental.
{While the algorithm in \cite{31} was effective for steady-state signals and frequency shifts, it showed significant errors when applied to dynamic signals with transient changes. Recognizing these limitations, a transient detection mechanism was introduced to segment the signal into time windows for more accurate power or energy calculation. Additionally, noise testing was conducted to evaluate its performance under realistic SNR in power systems, improving its robustness and accuracy in dynamic and noisy conditions.}

As a reminder, the remaining sections are organized as follows: The principle of the proposed CSEWT algorithm is introduced in Section \ref{sec02}. Various of tests and results are described in Section \ref{sec03}. Finally, the conclusion is drawn in Section \ref{sec04}.

\section{Principle of the proposed CSEWT algorithm}
\label{sec02}
\subsection{General process}

\begin{figure}[tp!]
	\centering
	\includegraphics[width=0.475\textwidth]{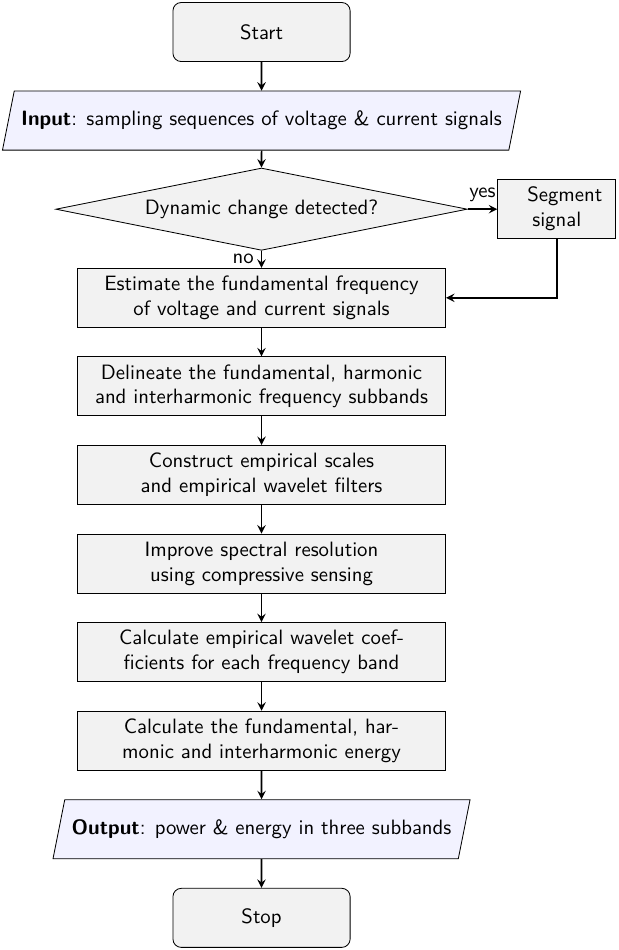}
	\caption{The general metering process of the proposed CSEWT.}
	\label{Fig.1}
\end{figure}

Fig.~\ref{Fig.1} illustrates the overall metering process of the proposed CSEWT. The first step involves signal pre-processing: voltage and current signals are sampled using a sampling frequency $f_s$ and a time window length of 200\,ms (see \cite{30}). 
The first step involves data pre-processing to precisely estimate the fundamental frequency. A transient detection is conducted to ensure that the proposed algorithm is effective for dynamic voltage and current signals.
Next, the fundamental frequency within the sampling window is estimated. Based on this estimate, frequency subbands corresponding to the fundamental, harmonic, and interharmonic components are defined. Empirical scale and wavelet filters are then constructed for these subbands. To enhance frequency resolution and align with the filters, the signal spectrum is subdivided using CS.
Following this, the signal spectrum is multiplied by the filter coefficients, and an inverse Fourier transform is applied to obtain the EWT coefficients for each frequency subband. Finally, the electric energy components associated with the fundamental, harmonic, and interharmonic frequencies are calculated. The following subsections provide a detailed explanation of each of these steps.

\subsection{Signal pre-processing}
The pre-processing step aims to accurately estimate the fundamental frequency. To enhance estimation precision, transient detection is incorporated, as transients can shift fundamental frequency estimates. Transient disturbances typically introduce high-frequency components into the signal. Thus, detecting high-frequency components helps identify whether a transient disturbance has occurred. In the proposed algorithm, frequencies are divided into low and high bands using $f_{s}/4$ as the boundary. If the amplitude of components in the high-frequency band exceeds 3\% of the voltage or current signal, a transient is considered to have occurred. When a transient is detected, its location is used to segment the voltage and current sequences before estimating the fundamental frequency. {The location of a transient refers to the specific time point within the observation window where the high-frequency amplitude exceeds the threshold. This time point is used to segment the signal for more accurate analysis.} If no transient is detected, the fundamental frequency is estimated directly from the entire sequence.

{It is acknowledged that signals originating from non-transient processes, such as those generated by electronic converters, may inherently contain real high-frequency components. If the amplitude of this high-frequency components don't exceed the threshold value, no segmentation is performed, and the fundamental frequency is estimated directly from the entire sequence. If the threshold is exceeded, segmentation occurs, but the fundamental frequencies of the segmented windows are identical, ensuring no additional measurement error, with acceptable time limits. It should be noted that the resolved transients in this paper include: amplitude swell, harmonic disappearance and frequency shifts. As a result, we have addressed the potential limitations on CSEWT performance by ensuring that transient detection and segmentation allow accurate energy and power calculations, even in cases involving amplitude swell, harmonic disappearance, and frequency shifts. The algorithm maintains computational efficiency, with the added segmentation time remaining within acceptable limits.}

\subsection{Full-band decomposition}
\label{sub:decompos}
After determining the fundamental frequency of the signal, $f_1$, the frequency subbands are defined as follows.
\begin{eqnarray}
\left[0,f_1-\frac{f_1}{k}\right]\cup\left[f_1-\frac{f_1}{k},f_1+\frac{f_1}{k}\right]\cup\left[f_1+\frac{f_1}{k},2f_1-\frac{f_1}{k}\right]\nonumber \\ 
\cup\left[2f_1-\frac{f_1}{k},2f_1+\frac{f_1}{k}\right]\cup\left[2f_1+\frac{f_1}{k},3f_1-\frac{f_1}{k}\right]\cup\cdots,
\label{eq:1}
\end{eqnarray}
where $k$ is the number of fundamental cycles within the measurement time window. Note that $k$ can be either an integer or a non-integer. Asynchronous sampling during signal pre-processing occurs only in the presence of frequency offsets or transient disturbances—such as voltage sags, swells, interruptions, and transient pulses—within the measurement window. These events may partition the time window into sub-windows, resulting in non-integer values of $k$.

\begin{figure}[tp!]
	\centering
	\includegraphics[width=0.5\textwidth]{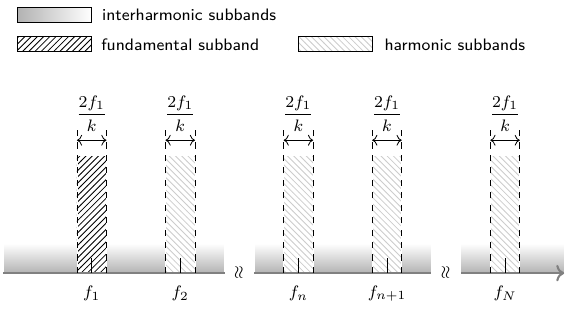}
	\caption{Division of frequency subbands. Three categories of frequency bands are classified: the fundamental subband, the harmonic subbands, and the interharmonic subbands. The width of the fundamental and each harmonic subband is $2f_1/k$. }
	\label{Fig.2}
\end{figure}

The frequency components near the fundamental frequency are defined as the fundamental frequency subband, specifically $[f_1-\frac{f_1}{k},f_1+\frac{f_1}{k}]$. Similarly, the frequency components near the harmonics are referred to as harmonic frequency subbands, expressed as  $[f_n-\frac{f_1}{k},f_n+\frac{f_1}{k}]$ where $n=2,3,4,...$. The remaining signal components are classified into interharmonic frequency subbands. These three types of subbands—fundamental, harmonic, and interharmonic—are illustrated in Fig.~\ref{Fig.2}. 

An orthogonality or quasi-orthogonality condition among different frequency subbands can greatly simplify power measurement. In the subband division presented, it is well known that, the fundamental frequency and its harmonics, as well as any two harmonic components, are mutually orthogonal. Active power is only generated between two components within the same frequency subband. In fact, this principle can be extended to all subbands. As demonstrated in \cite{32}, any cross-term generated between two frequency subbands becomes negligible when their center frequencies are separated by more than $f_1/k$ (in this case, 5 Hz), providing an approximation of the orthogonality condition.

\subsection{Filter bank construction using EWT}
\label{sub:EWT}
To extract the eigenvalues for fundamental, harmonic, and interharmonic frequency bands, it is essential to design appropriate filter banks. For this purpose, the EWT has been utilized. {EWT is adaptive and can generate the best wavelet window according to the segmentation of the signal spectrum, which is especially suitable for the analysis of dynamic signals.} Originally proposed by Jerome Gilles in \cite{33}, EWT defines each frequency band, including the transition bands, as a set of low-pass and band-pass filters on a compactly supported Fourier axis. The Fourier axis is normalized to simplify the division and segmentation process.

\begin{figure}[tp!]
	\centering
	\includegraphics[width=0.5\textwidth]{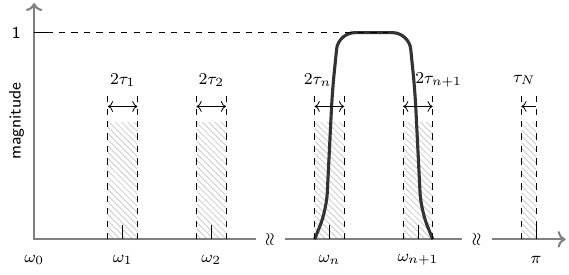}
	\caption{Schematic diagram of the segmentation of the Fourier axis. The width of the transition band is $2\tau_{n}$ ($2\gamma \omega_{n}$). The midpoint of the transition zone is $\omega_{n}$.}
	\label{Fig.3}
\end{figure}

Based on Shannon's criterion, the normalized segmented spectrum within the range of [0,$\pi$] of the signal to be measured is presented in Fig. \ref{Fig.3}.
Assuming that the number of amplitude-modulated frequency-modulated (AM-FM) components is $N$, it forms $N+1$ boundaries, i.e. $\omega_0,~\omega_1,~\omega_2,...,\omega_N$, and $N$ segments $\Lambda_n=[\omega_{n-1},\omega_{n}],n=1,2,...,N$.
In Fig.\ref{Fig.3}, $ 2\tau_n=2\gamma \omega_{n}(0 \textless \gamma \textless 1) $ denotes the width of the transition zone between two segments. To obtain the boundary condition, it defines that at the origin $\omega_{0}=0$ and the final point $\omega_{N}=\pi$. For the remaining $N-1$ boundaries, the extreme points of the Fourier spectrum are ranked in descending order, with the first $N-1$ extreme values retained (If insufficient, be supplemented with zero), and
$\omega_n$ is determined as the midpoint between two consecutive extreme points $\Omega_{n}$, i.e.,
\begin{eqnarray}		
\omega_{n}&=& (\Omega_n + \Omega_{n+1})/2,~~n = 1,2,...,N-1.
	\label{eq:2}
\end{eqnarray}

EWT defines each empirical wavelet of bandwidth $ \Lambda_n $ as a bandpass filter. Drawing on the Littlewood-Paley and Meyer wavelets, the empirical scale function $ \hat {\varphi_n}(\omega) $ and empirical wavelet function $ \hat {\psi_n}(\omega) $ for $ n \geq 0 $ are expressed as
\begin{equation}
		\hat {\varphi_n}(\omega) =
		\begin{cases}
			1, \hspace{1cm} \mathrm{if}~\left| \omega \right| \leq (1-\gamma)\omega_n \\~
			\\ \cos\left[\displaystyle\frac{\pi}{2} \beta \left(\displaystyle\frac{1}{2\gamma\omega_{n}}(\left| \omega \right| -(1-\gamma)\omega_{n})\right)\right], 
			\\ ~~\hspace{1cm}\mathrm{if}~(1-\gamma)\omega_{n} \leq \left| \omega \right| \leq (1+\gamma)\omega_{n} \\~
			\\ 0, \hspace{1cm} \mathrm{otherwise}
		\end{cases}
	\label{eq:3}
\end{equation}
\begin{equation}
		\hat {\psi_n}(\omega) =
		\begin{cases}
			1, \hspace{0.5cm}\mathrm{if}~(1+\gamma)\omega_{n} \leq \left| \omega \right| \leq (1-\gamma)\omega_{n+1} 
			\\ ~\\
   \cos\left[\displaystyle\frac{\pi}{2} \beta (\frac{1}{2\gamma\omega_{n+1}}(\left| \omega \right| -(1-\gamma)\omega_{n+1}))\right], 
			\\~ \hspace{0.5cm}\mathrm{if}~(1-\gamma)\omega_{n+1} \leq \left| \omega \right| \leq (1+\gamma)\omega_{n+1} 
			\\ ~\\
   \sin\left[\displaystyle\frac{\pi}{2} \beta (\frac{1}{2\gamma\omega_{n}}(\left| \omega \right| -(1-\gamma)\omega_{n}))\right], 
			\\~ \hspace{0.5cm} \mathrm{if}~(1-\gamma)\omega_{n} \leq \left| \omega \right| \leq (1+\gamma)\omega_{n} 
			\\ ~\\
   0, \hspace{0.5cm} \mathrm{otherwise}
		\end{cases}
  	\label{eq:4}
\end{equation}
Here $\beta(x)$ is a function satisfying 1) continuously differentiable on the interval $[0,1]$, 2) $\beta(x)=0$ when $x\leq0$; $\beta(x)=1$ when $x\geq1$, and 3) $\beta(x)+\beta(1-x)=1$ $\forall x \in [0,1]$. Here we take $\beta(x)=x^4(35-84x+70x^2-20x^3)$~\cite{33}. 

\begin{figure}[htp!]
	\centering
	\includegraphics[width=0.45\textwidth]{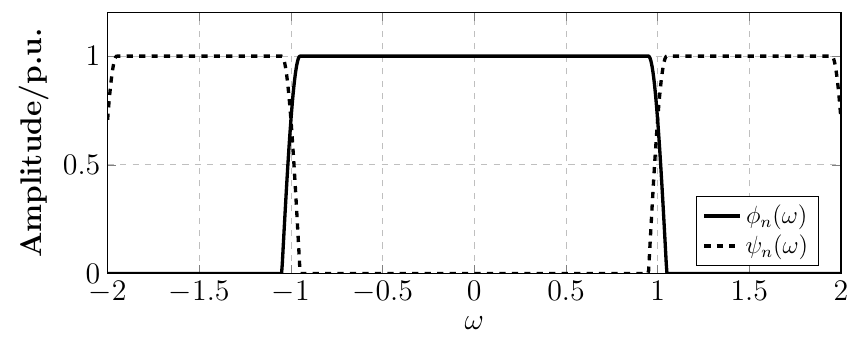}
	\caption{The transfer function of EWT ($\gamma=0.01$).}
	\label{Fig.4}
\end{figure}

{Wavelets are typically used overlapping, in that case, the effect from overlapping is reduced by adjusting $ \gamma < \min_n \left( \frac{\omega_{n+1}-\omega_n}{\omega_{n+1}+\omega_n} \right) $ to obtain tight support, making the measurement error smaller. The transfer function of EWT for $\gamma=0.01$ is shown in Fig.\ref{Fig.4}.} Following the above step, the filter bank on the normalized Fourier axis is constructed.

\subsection{Frequency resolution matching using CS}

For filter banks, insufficient spectral resolution leads to an overly wide transition band, making it difficult to accurately delineate discrete frequency subbands. In most cases, a spectral resolution of 0.5\,Hz is recommended for the filter bank. However, the spectral resolution of the voltage and current signals being measured is 5\,Hz, corresponding to a time window length of 200\,ms. As a result, it is not feasible to directly apply the filter bank (0.5\,Hz) to the voltage and current signals (5\,Hz). To ensure the filter bank obtained in \ref{sub:EWT} can be effectively used for feature extraction from the segmented spectrum signal described in \ref{sub:decompos}, it is necessary to match the spectral resolution of both the filter bank and the measured signals.
Since grid voltage and current signals consist of a finite number of spectral components, and the basis vectors of the FFT are orthogonal—satisfying the core principles of CS, namely sparsity and incoherence \cite{34}—CS is employed to enhance the frequency resolution of the voltage and current signals.
CS enables the reconstruction of the original signal that meets sparsity requirements using nonlinear reconstruction methods from a limited number of observations~\cite{35,36,37}. M. Betocco \textit{et al} combined CS with FFT to propose a highly accurate super-spectral resolution algorithm known as Compressive Sensing Discrete Fourier Transform (CSDFT). This algorithm has since been applied to address issues such as power quality monitoring, synchronized phasor estimation \cite{38}, and spectral analysis of signals \cite{39}. 

Using CS, the voltage and current spectrum can be represented using a multi-measurement vector as 
\begin{equation}
	\label{eq:5}
		S \approx D \alpha,
\end{equation}
where $ S $ is the observation vector, composed of the coefficients from the $N$-point DFT of the voltage and current signals, with a spectral resolution of $ \Delta f = {f_s}/{N} $. $ S $ is written as
\begin{equation}
	\label{eq:6}
		S =
		\begin{bmatrix}
			\hat{U}(k_1)& \hat{I}(k_1) \\ 
			\vdots & \vdots \\ 
			\hat{U}(k_N) & \hat{I}(k_N)
		\end{bmatrix}.
\end{equation}
$\hat{U}(k_n)$ and $\hat{I}(k_n)$ denote the frequency spectrum sequences of voltage and current, respectively, and $k_n$ ($n=1,2,...,N$) is the sequence index.

In (\ref{eq:5}), $D$ is the measurement matrix, composed of the Dirichlet kernel function, and is written as 
\begin{equation}
	\label{eq:7}
		D =
		\begin{bmatrix}
			\ D_{11} & \cdots & D_{1N^{\prime}} \\ 
			\vdots & \cdots & \vdots \\ 
			\ D_{N1} & \cdots & D_{NN^{\prime}}
		\end{bmatrix}, 
\end{equation}
with its elements characterized as
\begin{equation}
	\label{eq:8}
		D_{qr}=\frac{\sin\pi N\left(\displaystyle\frac{q}{N}-\frac{r}{N^{\prime}}\right)}{N \sin\pi \left(\displaystyle\frac{q}{N}-\frac{r}{N^{\prime}}\right)} e^{-\mathrm{j}\pi (N-1)(\frac{q}{N}-\frac{r}{N^{\prime}})}.
  \end{equation}
where $N^{\prime}=PN$ is the number of high-resolution spectral lines after refinement, {$1/N$ is a normalization factor,} and $P$ is the interpolation factor {($P$ is an integer)}. $q$ ($q=1,2,...,N$) is the index of the spectral line when the spectral resolution is $\Delta f$, and $r$ ($r=1,2,...,N^{\prime}$) is the index of the spectral line when the spectral resolution is $\Delta f^{\prime}$. 

$\alpha$ represents the vector of the signal to be refined, composed of refined spectral coefficients. Under this condition, the spectral resolution becomes $ \Delta f^{\prime} = {f_s}/{N^{\prime}} $. $ \alpha $ is written as 
\begin{equation}
	\label{eq:9}
		\alpha =
		\begin{bmatrix} 
			\hat{U}(\omega_1) & \hat{I}(\omega_1) \\ 
			\vdots & \vdots \\ 
			\hat{U}(\omega_{N^{\prime}}) & \hat{I}(\omega_{N^{\prime}}) 
		\end{bmatrix}.
\end{equation}
By making $N^{\prime}=10N$, $\alpha$ has a spectral resolution of 0.5 Hz and can then be operated with the filter bank.

\subsection{Calculation of wavelet coefficients}

Interharmonics are characterized by time-varying amplitude fluctuations and variable durations. If power calculations are conducted solely from a frequency-domain perspective using $UI\cos\phi_{u-i}$, errors may arise due to the averaging of interharmonic energy across the entire time window. To address this, the advantages of time-frequency analysis using the EWT can be utilized to obtain empirical wavelet coefficients. This is done by first performing an inverse Fourier transform on the product of the filter bank and the sparsified spectrum.
 
The voltage $ u(t) $ and current $ i(t) $ are both power-limited signals. By taking the inner product of the voltage and current signals with the empirical scale function $ \hat {\varphi_n}(\omega) $ and the empirical wavelet function $ \hat {\psi_n}(\omega) $, respectively, the corresponding empirical wavelet coefficients can be obtained as
\begin{eqnarray}
	\label{eq:10}
		d^1_u(t)&=&<u(t),\varphi_{1}(t)>= \rm{iFFT}(\hat{U}(\omega)\cdot\hat{\varphi_{1}}(\omega)),\\
	\label{eq:11}
		d^n_u(t) &=& <u(t),\psi_{n}(t)> = \rm{iFFT}(\hat{U}(\omega)\cdot\hat{\psi_{n}}(\omega)),\\
	\label{eq:12}
		d^1_i(t) &=& <i(t),\varphi_{1}(t)> = \rm{iFFT}(\hat{I}(\omega)\cdot\hat{\varphi_{1}}(\omega)),\\
	\label{eq:13}
		d^n_i(t) &=& <i(t),\psi_{n}(t)> = \rm{iFFT}(\hat{I}(\omega)\cdot\hat{\psi_{n}}(\omega)).
\end{eqnarray}
where, $ d^1_u(t) $ and $ d^1_i(t) $ are the empirical wavelet coefficients of $ u(t) $ and $ i(t) $ at the first frequency band (the lowest frequency band), while $ d^n_u(t) $ and $ d^n_i(t) $ represent the empirical wavelet coefficients at the nth frequency band(\textit{n}=2,3,...,$K$), $K$ is the number of frequency bands.
$ \hat{U}(\omega) $ and $ \hat{I}(\omega) $ are the Fourier transform of $ u(t) $ and $ i(t) $, respectively. $ \rm {iFFT}(\cdot) $ represents the inverse Fourier transform.

\subsection{Calculation of electrical energy}
After obtaining the empirical wavelet coefficients, the fundamental, harmonic, and interharmonic frequency components of the signal can be reconstructed. From these reconstructed components, the corresponding fundamental energy, harmonic energy, and interharmonic energy can then be calculated.

The reconstruction process is as follows: when an orthogonal wavelet basis is selected, the discrete voltage $ u(n) $ and current $ i(n) $ can be reconstructed as
\begin{eqnarray}
	\label{eq:14}
		u(n) &=&\sum_{s}d^{1}_u(s)\phi_{1}(n-s)+\sum_{p=2}^{K}\sum_{s}d^{p}_u(s)\psi_{p}(n-s) \nonumber\\
		&=&\sum_{s}d^{1}_u(s)\phi_{1,s}(n)+\sum_{p=2}^{K}\sum_{s}d^{p}_u(s)\psi_{p,s}(n).
\end{eqnarray}
\begin{eqnarray}
	\label{eq:15}
		i(n) &=&\sum_{s}d^{1}_i(s)\phi_{1}(n-s)+\sum_{p=2}^{K}\sum_{s}d^{p}_i(s)\psi_{p}(n-s) \nonumber\\
		&=&\sum_{s}d^{1}_i(s)\phi_{1,s}(n)+\sum_{p=2}^{K}\sum_{s}d^{p}_i(s)\psi_{p,s}(n).
\end{eqnarray}
where, $ d^{1}_u(s) $, $ d^{p}_u(s) $, $ d^{1}_i(s) $ and $ d^{p}_i(s) $ are the discrete empirical wavelet coefficients for voltage and current, respectively. $ \phi_{1,s}(n) $ and $ \psi_{p,s}(n) $ represent the basis functions of each mutually orthogonal subspace in the decomposition performed by EWT, {and have the following properties:

\begin{equation}
\label{eq:16}
\begin{aligned}
    \langle \phi_{1,s}(n), \phi_{1,s}(n) \rangle &= 1, \\
    \langle \phi_{1,s}(n), \psi_{p,s}(n) \rangle &= 0, \quad p \geq 2, \\
    \langle \psi_{p,s}(n), \psi_{p,s}(n) \rangle &= 1, \\
    \langle \psi_{p,s}(n), \psi_{p',s}(n) \rangle &= 0, \quad p \neq p'.
\end{aligned}
\end{equation}

Based on the orthogonality of the basis functions as (\ref{eq:16})}, {
both the EWT and CS approximately satisfy energy conservation when the time domain signal is transformed to the frequency domain for analysis. The analysis of signal power preservation in CSEWT is detailed in the appendix. This demonstrates that, despite inherent approximations, the proposed method maintains near energy conservation, ensuring its reliability for frequency-domain analysis of power system signals. Therefore,}
the fundamental energy, harmonic energy, and interharmonic energy are calculated, followed by
\begin{equation}
	\label{eq:17}
		W_p=\sum_{s}d^{p}_u(s)d^{p}_i(s)T_s.
\end{equation}

Based on the designed EWT frequency subbands, the fundamental frequency corresponds to the node $ p=2 $, which lies within the range 
$\left[f_1 - \frac{f_1}{k}, f_1 + \frac{f_1}{k} \right]$.
The remaining harmonics correspond to nodes $p=2h$, where $h=2,3,\dots,H$ and $H$ is the highest-order harmonic considered. Components at nodes where $p \neq 2h$ are classified as interharmonics, contributing to the interharmonic energy. As a result, the fundamental, harmonic, and interharmonic energies can all be calculated from this classification.

\section{Performance tests}
\label{sec03}
{This section assesses the performance of the proposed CSEWT algorithm under a comprehensive set of test conditions, including steady-state, amplitude-phase modulation, amplitude swell, harmonic and interharmonic disappearance, frequency shift (to account for potential deviations in the fundamental frequency post-transient events), and noise interference. These tests are specifically designed to ensure the algorithm's robustness and accuracy in measuring power under dynamic conditions, particularly in scenarios involving grid voltage and current signals with interharmonic components.}

For the first five tests, Gaussian white noise with a signal-to-noise ratio (SNR) of 60 dB was added to the voltage and current signals, while in the noise test, the algorithm's response to Gaussian white noise was evaluated across different SNRs (ranging from 40 dB to 80 dB). The FFT, EWT, and CSDFT methods were selected for comparison. FFT is the commonly used frequency domain power metering algorithm, while EWT is the appropriate time-frequency domain power metering algorithm, and CSDFT improves the spectral resolution of DFT using the CS technique. The sampling frequency was set to 6400\,Hz (128 points in one fundamental period), with a measurement duration of 10 fundamental cycles (0.2\,s). The evaluation metric used is the absolute value of the relative error of each frequency component, denoted as $E_{fc}$. Each test was repeated 100 times, and the mean value with a standard deviation was given as the output.

\subsection{Steady-state condition}
The voltage and current test signals are given as
\begin{eqnarray}
	\label{eq:18}
		s(t)&=&\sum\limits_{h=1}^{H}A_h\sin (2\pi f_ht+\phi_h) \nonumber \\
		&&+\sum\limits_{ih=1}^{N_{ih}}A_{ih}\sin (2\pi f_{ih}t+\phi_{ih})+s_{N}(t),
\end{eqnarray}
where the amplitude and frequency of the fundamental component are set as $A_1$ = 1\,p.u. and $f_1$ = 50\,Hz. A set of harmonics ($ h\in[2,9] $) is added in both the voltage and current signals, and the amplitude $A_h$ is set to 0.1\,p.u. with frequencies $f_h=hf_1$. 
Three interharmonic components, $f_{ih}$ of { 70\,Hz(multiples of 5\,Hz), 232.5\,Hz(multiples of 0.5\,Hz), 369\,Hz(multiples of 1\,Hz), which represent different levels of resolution to verify the proposed algorithm's excellent performance}, are included in the signal model, and the amplitude of each component is also set as $ A_{ih} $=0.1\,p.u.. $ s_{N}(t)$ is a 60 dB of Gaussian white noise.
In the test, the power factor (PF) of each component (fundamental, 2-9th harmonics, 3 interharmonics) is set to 0, 0.5L and 0.5C, where the phase difference between voltage and current signals is 0, $ \pi/3 $, and $ -\pi/3 $, respectively. 

\begin{figure}[tp!]
	\centering
	\includegraphics[width=0.475\textwidth]{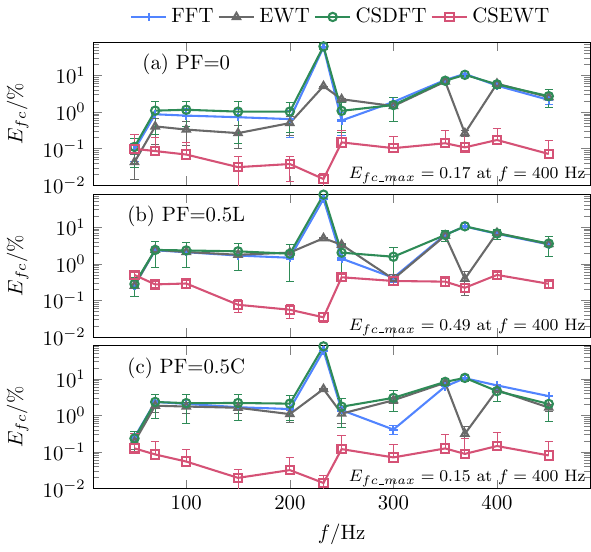}
	\caption{$E_{fc}$ of four algorithms under steady-state tests. From (a) to (c), the PF is set to 0, 0.5L and 0.5C, and $\phi_{u-i}$ is 0, $ \pi/3 $, and $ -\pi/3 $, correspondingly.}
	\label{Fig.5}
\end{figure}

Fig.~\ref{Fig.5} presents the relative error \( E_{fc} \) at different frequencies for PF of 0, 0.5L, and 0.5C, respectively. As is seen, variations in PF have minimal impact on the \( E_{fc} \) values across the four algorithms. Among them, CSEWT demonstrates a consistently lower \( E_{fc} \) with less fluctuation. In contrast, FFT and CSDFT show significant variability, with \( E_{fc} \) reaching notably high values at certain frequencies, such as 232.5\,Hz.
The FFT algorithm, with a spectral resolution of 5\,Hz, is unable to accurately extract the interharmonic component at 232.5\,Hz. Although the CSDFT algorithm improves spectral resolution, it also amplifies spectral leakage, resulting in inaccurate extraction of the interharmonic component at 232.5\,Hz. Therefore, both FFT and CSDFT are unsuitable for signals containing interharmonic components under steady-state conditions. On the other hand, EWT and CSEWT perform more effectively at different frequencies and are overall better suited to handling both harmonic and interharmonic components.

\subsection{Amplitude-phase modulation}

Note that in power systems, voltage signals typically exhibit much less variability compared to current signals. Therefore, in this test, only the current signal is modulated in terms of amplitude, phase, and frequency. {The inherent calculus relationship between frequency and phase establishes a strong theoretical equivalence between frequency modulation (FM) and phase modulation (PM). Specifically, the instantaneous frequency is the time derivative of the instantaneous phase, and conversely, the phase is the integral of the frequency. Under sinusoidal modulation, FM and PM can produce nearly identical signals, differing only by a constant phase offset. Consequently, this work employs amplitude-phase modulation as a unified framework, reducing complexity by leveraging the interchangeability of FM and PM where applicable.}

The voltage signal remains in the form given by (\ref{eq:18}), while the current signal is defined as
\begin{eqnarray}
	\label{eq:19}
		i(t) &=&\sum\limits_{h=1}^{H} \left[  
		\begin{array}{l}
		A_h(1+k_{x} \sin (2\pi t)) \nonumber\\
		\cdot \sin (2\pi  f_ht+\varphi_h + k_{a} \sin (2\pi t)) 
		\end{array}
		\right] \\
		&&+\sum\limits_{ih=1}^{N_{ih}}A_{ih}\sin (2\pi f_{ih}t+\varphi_{ih})+s_{N}(t),
\end{eqnarray}
where $ k_{x} $ represents the amplitude modulation coefficient, and $ k_{a} $ is the phase modulation coefficient. In the modulation test, $k_{x}$ is set as 0.1 and $ k_{a} $ is set as 0.4. 

\begin{figure}[tp!]
	\centering
	\includegraphics[width=0.475\textwidth]{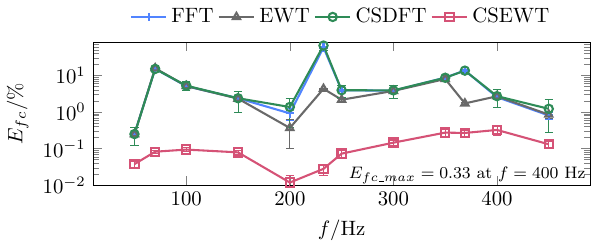}
	\caption{$E_{fc}$ of four algorithms under a modulation test. In this example, $k_{x}= 0.1$ and $k_{a} = 0.4$.}
	\label{Fig.6}
\end{figure}

The results of the modulation test are shown in Fig.~\ref{Fig.6}. In the modulation, the signal generates additional interharmonic components, highlighting the limitations of the FFT and CSDFT algorithms, which are unable to accurately extract interharmonic features. As a result, both FFT and CSDFT exhibit higher \( E_{fc} \), with a significant increase in \( E_{fc} \) near interharmonic frequency points.
In contrast, EWT and CSEWT calculate the electrical energy of the components based on frequency subbands, leading to lower \( E_{fc} \) values for each frequency component. However, the limited spectral resolution of the EWT makes it challenging to accurately capture some variations, especially at interharmonic positions. CSEWT, on the other hand, demonstrates a clear advantage over the other three algorithms in terms of performance.

It is also important to note that variations in \( k_x \) and \( k_a \) had minimal impact on the \( E_{fc} \) trends across all four algorithms.

\subsection{Amplitude swell}
A swell is an increase in amplitude above 1.1 p.u. for a duration over 0.5 cycle \cite{40}. Here for the swell test, the amplitudes of voltage and current jump from 1 p.u. to 1.4 p.u. at a specific time $t_a=0.115$\,s. 

\begin{figure}[tp!]
	\centering
	\includegraphics[width=0.475\textwidth]{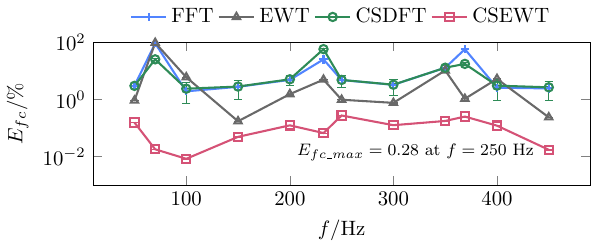}
	\caption{$E_{fc}$ of four algorithms under amplitude swell.}
	\label{Fig.7}
\end{figure}

The test result is shown in Fig. \ref{Fig.7}. The \( E_{fc} \) of FFT and CSDFT, similar to the performance of other tests, is high at interharmonic frequencies. The EWT, which measures electrical energy based on frequency subbands, maintains a lower \( E_{fc} \) expect for the interharmonic component $f=70\,$Hz. This is because the spectral resolution of EWT is limited at 5\,Hz, and when the detection signal is at the subband boundary, it becomes insufficient to accurately measure each component, thereby leading to an increase in \( E_{fc} \). While, the CSEWT can well remain a low \( E_{fc} \) in the full frequency band when the amplitude swells. 

A similar result is obtained when varying the swell amplitude or the time triggering the event ($t_a$).

\subsection{Harmonic and interharmonic disappearance}
Adhering to IEC 61000-4-7 \cite{30}, power devices such as electronic frequency converters can cause significant harmonic and interharmonic amplitude fluctuations, including the abrupt disappearance of these components. Therefore, in this test, the harmonic and interharmonic components of voltage and current disappear at a specific time, $t_b$. The test signals are defined the same as the steady-state test in (\ref{eq:18}) when $t<t_b$. The harmonic and interharmonic components disappear at $ t_b=0.13 $\,s, leaving only the fundamental component with $t\geq t_b$.

\begin{figure}[tp!]
	\centering
	\includegraphics[width=0.475\textwidth]{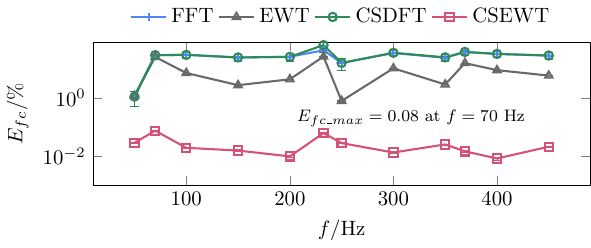}
	\caption{$E_{fc}$ of four algorithms under harmonic disapprearance.}
	\label{Fig.8}
\end{figure}

As shown in Fig.~\ref{Fig.8}, CSEWT demonstrates significantly higher measurement accuracy compared to the other three algorithms. This superiority stems from the fact that FFT and CSDFT are frequency-domain algorithms. {So FFT and CSDFT redistribute the energy of harmonic and interharmonic components across the entire observation window when these components partially vanish. This redistribution causes waveform distortion, leading to reduced amplitude and extended duration in the reconstructed signals. Consequently, the distortion introduces measurement errors in electrical energy. Due to this limitation,} FFT and CSDFT exhibit a higher overall \( E_{fc} \) compared to EWT and CSEWT. For the EWT, the low spectral resolution hinders the accurate extraction of interharmonic frequency components, resulting also a low accuracy. Notably, at 70\,Hz, the \( E_{fc} \) of EWT exceeds that of both FFT and CSDFT. 
In summary, CSEWT is particularly well-suited for scenarios involving the disappearance of harmonic components.

\subsection{Frequency shift}

\begin{figure*}[tp!]
	\centering
	\includegraphics[width=0.9\textwidth]{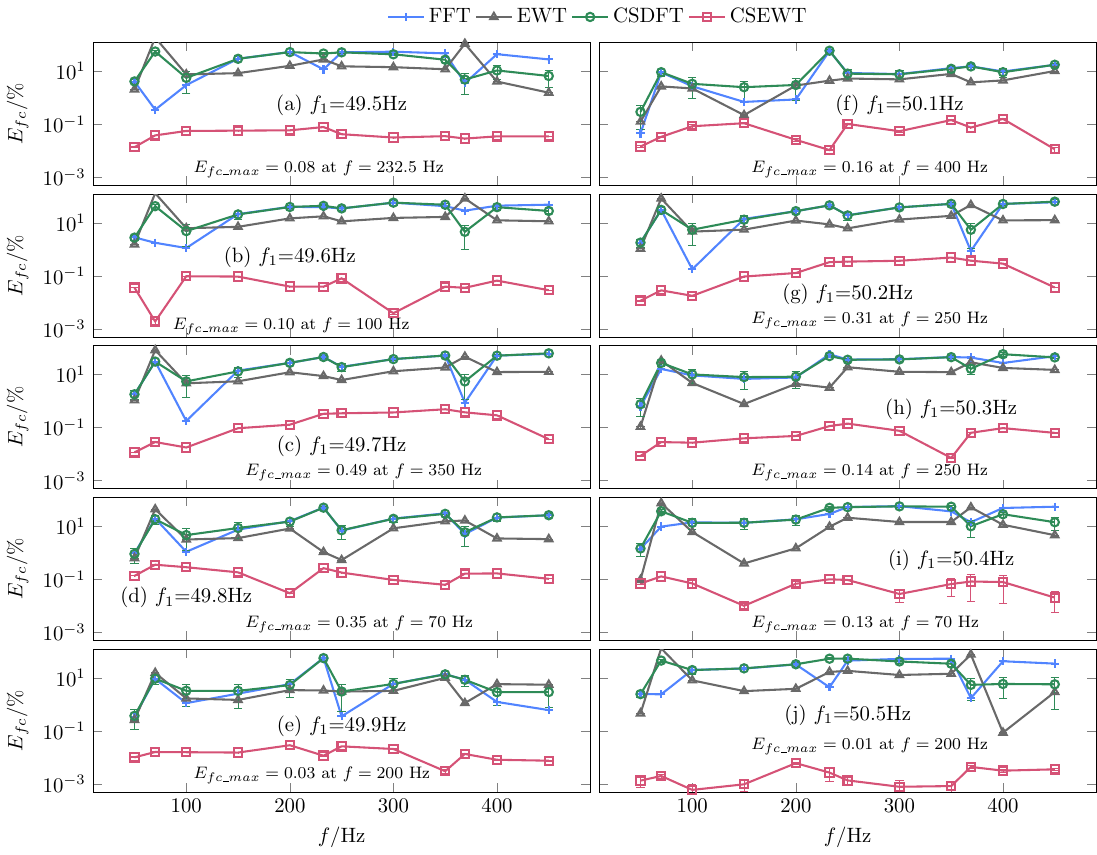}
	\caption{$E_{fc}$ of four algorithms under frequency shift. $f_{1}$ increases from 49.5\,Hz to 50.5\,Hz with a step of 0.1\,Hz.}
	\label{Fig.9}
\end{figure*}

In the frequency shift test, the voltage and current signals defined in (\ref{eq:18}) are employed but with the fundamental frequency $f_1$ ramping from 49.5\,Hz to 50.5\,Hz in steps of 0.1\,Hz(\( f_1 \)=50\,Hz is equivalent to case \textit{A}, and is not shown.). All other parameter settings remain unchanged, consistent with case \textit{A}. It should be noted that when the fundamental frequency $f_1$ is shifted, the harmonic frequencies correspond to $hf_1$.

The test results are shown in Fig.~\ref{Fig.9}. The four algorithms exhibit different performances as \( f_1 \) increases from 49.5\,Hz to 50.5\,Hz(without 50\,Hz). The \( E_{fc} \) of the FFT and CSDFT algorithms show similar trends, with a pronounced error peak at 232.5\,Hz (c,d,e,f,g,h) and substantial fluctuations in the range [350\,Hz,400\,Hz] (a,b,i,j), particularly for CSDFT in the corresponding cases. EWT performs poorly at certain frequency points, especially at 70\,Hz and 369\,Hz, where \( E_{fc} \) values are significant (a,b,c,d,g,h,i,j); however, at other frequency points, the \( E_{fc} \) is relatively small. CSEWT excels in the frequency shift test, demonstrating optimal and consistent performance with the lowest \( E_{fc} \) when \( f_1 \) varies within the range [49.5\,Hz, 50.5\,Hz].

It is worth noting that FFT, EWT, and CSDFT sometimes exhibit lower \( E_{fc} \) at specific frequency points. For instance, FFT shows lower errors at 70\,Hz (a,f,j), CSDFT at 369\,Hz (b,c,d,e,f,j), and EWT at 150\,Hz (d,e,f,g,h,i). This phenomenon can be attributed to the shift in the fundamental frequency, which leads to a corresponding shift in harmonic frequencies. Notably, the degree of shift increases with harmonic order. In (i) \( f_1 = 50.4\,\text{Hz} \), when \( f_1 \) shifts by 0.4\,Hz, the highest harmonic in the original signal is shifted by 3.6\,Hz. This shift results in a more complex positional relationship between harmonics and interharmonics. The accuracy of FFT and CSDFT may improve if harmonics move further away from the interharmonic positions than before the shift in \( f_1 \), and conversely, accuracy may decrease if they move closer. Given that the original signal contains three interharmonic components, the positions of harmonics and interharmonics vary non-monotonically, leading to greater volatility in the accuracy of FFT and CSDFT. There are even instances where the trends of these two variations are not identical at specific frequency points, such as 232.5\,Hz in (j) \( f_1 = 50.5\,\text{Hz} \).

The large errors observed in EWT are primarily due to its lack of frequency resolution. Most of the original harmonic frequency components no longer align with the 5\,Hz interval points after the shift in \( f_1 \). The significant errors in some cases (a,b,c,g,i,j) occur because, when \( f_1 \) shifts, the components extracted by EWT no longer belong to the original groupings.

In conclusion, CSEWT is an effective approach for power metering when frequency shift issues are considered.

\subsection{Noise test}
In the noise test, the voltage and current test signals are defined the same as the signal used in the harmonic disappearance test, and the difference is changing the noise level $ s_{N}(t) $. Based on the characterization of voltage or current signals from the power system, the signal-to-noise ratio (SNR) is set to 40, 50, 60, 70, and 80 dB, respectively.

\begin{figure}[tp!]
	\centering
	\includegraphics[width=0.475\textwidth]{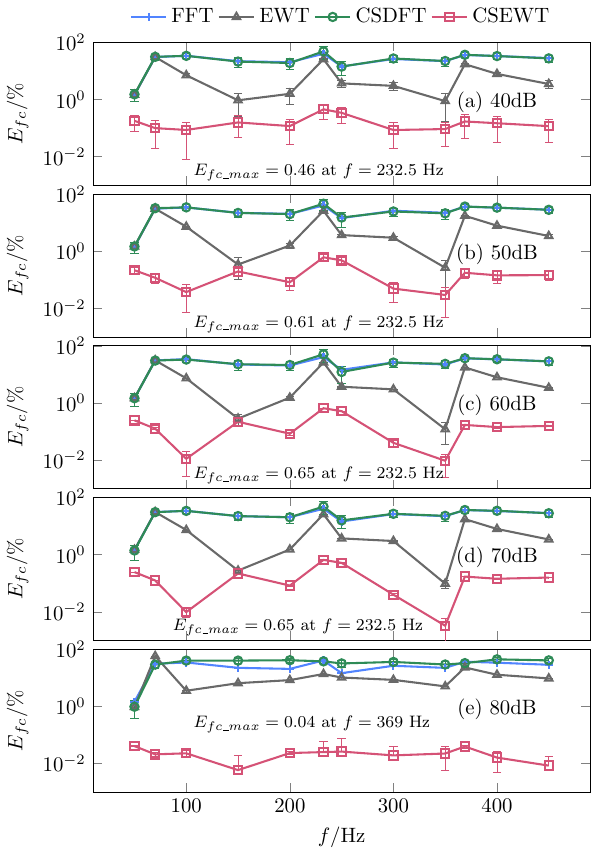}
	\caption{$E_{fc}$ of four algorithms under noise test. SNR increases from 40\,dB to 80\,dB with a step of 10\,dB.}
	\label{Fig.10}
\end{figure}

As shown in Fig.~\ref{Fig.10}, FFT, EWT, and CSEWT exhibit good noise immunity stability when the SNR lies within the range of [40\,dB, 80\,dB]. In contrast, CSDFT is more susceptible to noise interference, as evidenced by the reduction in the range of \( E_{fc} \) fluctuations when the SNR is increased to 80\,dB.

In general, FFT and CSDFT, both belonging to the frequency-domain class, exhibit a comparable pattern of \( E_{fc} \) variation and show larger errors compared to EWT and CSEWT, which belong to the time-frequency domain class. Compared to CSEWT, EWT shows a larger \( E_{fc} \) at interharmonic frequency points (70\,Hz, 232.5\,Hz, and 369\,Hz). However, the \( E_{fc} \) of EWT is observed to be lower than that of CSEWT at 150\,Hz. 

Under low-SNR conditions, noise components can generate new frequency components that cannot be ignored. Due to the improved spectral resolution of CSEWT, this additional energy is captured, leading to increased measurement errors. In contrast, EWT focuses only on the original frequency components, which are less affected by noise, resulting in better accuracy. As the SNR increases, i.e., as noise decreases, the advantages of CSEWT become more pronounced. In the noise test, the \( E_{fc} \) of CSEWT remains consistently low and stable, indicating its robust performance under various noise conditions.

\subsection{Computational time analysis}
In the above six tests, all algorithms were implemented in Matlab 2020a and tested on a machine equipped with an Intel i7-10700K CPU and 32GB RAM. The execution times were measured using the built-in time module and averaged over 100 runs to mitigate variability.

It is known that the theoretical time complexity of FFT is $\mathcal{O}(N\log N)$ \cite{41}, where $N$ is the signal length. EWT consists of two steps, spectral analysis and adaptive filtering, with a time complexity of $\mathcal{O}(N\log N)+\mathcal{O}(KN)$ \cite{33}, where $K$ is the number of frequency bands. Theoretically, the time complexity of CSDFT or CSEWT can be approximately estimated as the sum of the time complexity of the CS and DFT or EWT, respectively. The time complexity of CS consists of three parts \cite{35}: 1) encode time: the process of generating a compressed measurement from the original spectral sequence, 2) sparsity update time: the duration of the iteration during which the sparse structure of the signal is updated, and 3) approximate recovery time: the time required to reconstruct the original signal from the compression measurement. The time complexity of the 
CS can be approximately estimated as $\mathcal{O}(M^3) + \mathcal{O}(T(N\log N + M^3))$, where $M$ is the size of the measurement matrix, and $T$ is the number of iterations. 
It should be noted that the time required for the CS process is not the same for CSDFT and CSEWT, because CSDFT requires iterative operations for each peak frequency point and 5 frequency points to the left and right (11 points in total, recommended by \cite{30}), while CSEWT only requires iterative operations on the frequency bands (already delimited) in which the peak frequency point is located.

The average execution times of four algorithms across the six tests are summarized in Table \ref{tab:1}. As shown, both FFT and EWT demonstrate rapid data processing capabilities, with average execution times significantly below 0.2\,s. In contrast, CSDFT exhibits a much longer processing duration, exceeding 0.2\,s, rendering it unsuitable for real-time applications. The computation time for CSEWT varies depending on the specific scenario. Specifically, for cases A and B, the computation time is under 0.1\,s, whereas for cases C through F, it increases moderately, ranging from 0.14\,s to 0.2\,s. This elevated computation time can be attributed to the algorithm's need to segment the signal and separate the frequency bands when processing dynamic signals. Such iterative signal processing, unique to CSEWT, results in a 40\% to about 100\% increase in computation time compared to the steady-state scenario.

\begin{table}[tp!]
    \centering
    \caption{Average time of four algorithms in six tests.}
    \label{tab:1}
    \begin{tabular}{c|c|c}
    \hline
    \hline
    \textbf{Case} & \textbf{Algorithms} & \textbf{Time (s)} \\ 
    \hline
    \multirow{4}{*}{\makecell{A\\ Steady-state condition}} 
                    & FFT   & 0.002 \\ \cline{2-3} 
                    & EWT   & 0.035 \\ \cline{2-3} 
                    & CSDFT & 0.265 \\ \cline{2-3} 
                    & CSEWT & 0.092 \\ \hline
    \multirow{4}{*}{\makecell{B\\ Amplitude-phase modulation}} 
                    & FFT   & 0.002 \\ \cline{2-3} 
                    & EWT   & 0.035 \\ \cline{2-3} 
                    & CSDFT & 0.266 \\ \cline{2-3} 
                    & CSEWT & 0.090 \\ \hline
    \multirow{4}{*}{\makecell{C\\ Amplitude swell}} 
                    & FFT   & 0.002 \\ \cline{2-3} 
                    & EWT   & 0.034 \\ \cline{2-3} 
                    & CSDFT & 0.269 \\ \cline{2-3} 
                    & CSEWT & 0.150 \\ \hline
    \multirow{4}{*}{\makecell{D\\ Harmonic and interharmonic\\ disappearance}} 
                    & FFT   & 0.002 \\ \cline{2-3} 
                    & EWT   & 0.034 \\ \cline{2-3} 
                    & CSDFT & 0.271 \\ \cline{2-3} 
                    & CSEWT & 0.195 \\ \hline
    \multirow{4}{*}{\makecell{E\\ Frequency shift}} 
                    & FFT   & 0.002 \\ \cline{2-3} 
                    & EWT   & 0.034 \\ \cline{2-3} 
                    & CSDFT & 0.270 \\ \cline{2-3} 
                    & CSEWT & 0.146 \\ \hline
    \multirow{4}{*}{\makecell{F\\ Noise test}} 
                    & FFT   & 0.002 \\ \cline{2-3} 
                    & EWT   & 0.034 \\ \cline{2-3} 
                    & CSDFT & 0.269 \\ \cline{2-3} 
                    & CSEWT & 0.197 \\ 
                    \hline
                    \hline
    \end{tabular}
\end{table}

\section{Conclusion}
\label{sec04}

{The non-stationary nature of grid voltage and current signals presents challenges for existing methods of electrical power measurement, making it difficult to accurately quantify fundamental, harmonic, and interharmonic components of power.} To address this issue, this paper proposes a new algorithm designed to enhance power measurement accuracy under non-stationary conditions, particularly in the presence of interharmonics. The proposed algorithm separates the fundamental, harmonic, and interharmonic components of the signal using a full-band decomposition method. Based on the empirical wavelet variation method, scale and wavelet functions with sufficiently high spectral resolution are designed, and a compressive sensing algorithm is employed to further improve the spectral resolution. The time-domain information of the empirical wavelet coefficients is then obtained via the inverse Fourier transform (IFFT), allowing for the accurate calculation of fundamental, harmonic, and interharmonic active electrical energy. {Comparative tests with FFT, EWT, and CSDFT under steady-state signals, dynamic signals, and noise conditions demonstrate that CSEWT offers three distinct advantages: 1) Under steady-state conditions, CSEWT achieves a measurement error of less than 4.94\%, compared to 7.82\% for EWT, 59.04\% for FFT and 77.33\% for CSDFT, fully meeting the error requirements for each frequency component. 2) Under dynamic conditions, CSEWT provides more accurate and stable measurements, with the measurement error reduced by about 1-2 orders of magnitude compared to FFT, EWT and CSDFT, demonstrating its superior stability. 3) Under noise conditions, CSEWT exhibits strong noise immunity, maintaining accurate measurement of each frequency component even at an SNR as low as 40\,dB, whereas FFT, EWT and CSDFT begin to show large fluctuations below 60\,dB.}

Future work will focus on exploring the potential application of the proposed algorithms in metering devices, such as standard power meters, particularly in scenarios involving interharmonic components. {Additionally, the study will incorporate uncertainty analysis once suitable hardware platforms become available.}

\section*{Acknowledgement}
The authors would like to thank graduates of our group, Dr. Yiqing Yu and Dr. Dongfang Zhao for valuable discussions on some technical issues.

{\appendix
\label{appendix}
The core components of the CSEWT include the Empirical Wavelet Transform (EWT), the Fast Fourier Transform (FFT), and Compressed Sensing (CS). The preservation of signal power within these three aspects is discussed as follows.

\begin{enumerate}
    \item \textbf{EWT}. The potential factors undermining energy conservation in EWT can be attributed to three primary aspects: non-ideal characteristics of the filter bank, imprecise band segmentation, and spectral leakage caused by filter transition bands. To mitigate these limitations, this study introduces enhanced filter designs featuring improved orthogonality and reduced transition bandwidths. These refinements effectively minimize inter-component energy leakage while ensuring that the segmentation boundaries align precisely with the spectral characteristics of the analyzed signals. As a result, the proposed algorithm achieves near-perfect energy conservation, as evidenced by the following relation:

\[
\|x(t)\|_2^2 \approx \sum_j \|x_j(t)\|_2^2.
\]

This demonstrates the algorithm's ability to preserve energy across the decomposed components, validating its efficacy in addressing the identified challenges.

\item \textbf{FFT}. FFT adheres to the principle of energy conservation, a concept formally articulated through Parseval's theorem:

\[
\int_{-\infty}^\infty |x(t)|^2 \, dt = \int_{-\infty}^\infty |X(f)|^2 \, df.
\]

This fundamental relationship ensures the preservation of signal energy during the transformation process, establishing an exact equivalence between the energy computed in the time domain and its representation in the frequency domain. By guaranteeing this equivalence, Parseval's theorem ensures that no energy dissipation occurs during the FFT, thereby maintaining the integrity of the signal's energy throughout the transformation.

\item \textbf{CS}. In CS, the energy characteristics of signals can be analyzed from two perspectives:

\textit{a) Energy relationship between compressed and original signals}:  
The compressed signal \(\mathbf{y}\) is obtained through linear measurement \(\mathbf{y} = \mathbf{\Phi x}\), where \(\mathbf{\Phi}\) represents the measurement matrix. Due to the dimensionality reduction from \(\mathbf{x} \in \mathbf{R}^N\) to \(\mathbf{y} \in \mathbf{R}^M\) (where \(M < N\)), the energy of the compressed signal is generally reduced:

\[
\|\mathbf{y}\|_2^2 = \|\mathbf{\Phi} \mathbf{x}\|_2^2 \leq \|\mathbf{x}\|_2^2.
\]

Equality holds only when \(\mathbf{\Phi}\) is energy-preserving, such as in the case of an orthogonal matrix.

\textit{b) Energy relationship between reconstructed and original signals}:  
The reconstructed signal \(\hat{\mathbf{x}}\) is obtained through optimization algorithms. Under ideal conditions, where the measurement matrix satisfies the Restricted Isometry Property (RIP), the energy relationship approximates:

\[
\|\hat{\mathbf{x}}\|_2^2 \approx \|\mathbf{x}\|_2^2.
\]

However, practical factors such as noise and reconstruction inaccuracies typically prevent exact energy conservation.
The proposed algorithm employs a Dirichlet kernel-based measurement matrix derived from the Fourier transform:

\begin{eqnarray*}
    X(k) &=& \sum_h \left[ A_h e^{j\theta_h} D_N\left(\frac{k}{N} - \frac{f_h}{f_s}\right)\right.\\ 
    &&\left.+ A_h e^{-j\theta_h} D_N\left(\frac{k}{N} + \frac{f_h}{f_s}\right) \right].
\end{eqnarray*}

This matrix approximately satisfies the RIP condition:

\[
(1 - \delta_k) \|\mathbf{x}\|_2^2 \leq \|\mathbf{\Phi} \mathbf{x}\|_2^2 \leq (1 + \delta_k) \|\mathbf{x}\|_2^2.
\]

For power system signals, where voltage and current components exhibit sparsity and relatively low noise, combined with the Orthogonal Matching Pursuit (OMP) reconstruction algorithm, the proposed method achieves near energy conservation. This ensures that the energy of the reconstructed signal closely approximates that of the original signal, making the approach particularly suitable for applications in power system analysis.

\end{enumerate}

In summary, while the FFT process maintains exact energy conservation, both the EWT and CS processes introduce some degree of energy loss. However, by employing optimized measurement matrices and advanced reconstruction algorithms, the proposed method achieves approximate energy conservation, making it well-suited for power system signal analysis. This balance between theoretical precision and practical applicability ensures the method's effectiveness in handling real-world power system signals while minimizing energy discrepancies.

}

\end{document}